\documentclass[%
reprint,
amsmath,amssymb,
aps,
]{revtex4-2}

\usepackage{amsmath}
\usepackage{graphicx}% Include figure files
\usepackage{dcolumn}% Align table columns on decimal point
\usepackage{bm}% bold math
\usepackage{tikz}
\usetikzlibrary{calc}
\usepackage{hyperref}% add hypertext capabilities
\definecolor{linkcolor}{HTML}{0000FF} % цвет ссылок
\definecolor{citecolor}{HTML}{0000FF} % цвет гиперссылок
\hypersetup{pdfstartview=FitH,  linkcolor=linkcolor, urlcolor=citecolor, citecolor=citecolor, colorlinks=true}
\begin{document}

\preprint{APS/123-QED}

\title{Architecture agnostic algorithm for reconfigurable optical interferometer programming}% Force line breaks with \\

\author{Sergei Kuzmin}
\author{Ivan Dyakonov}
\email{dyakonov@quantum.msu.ru}
\author{Sergei Kulik}
\affiliation{Quantum Technology Centre, Faculty of Physics, Lomonosov Moscow State University, Moscow, Russian Federation}
\date{\today}% It is always \today, today,

\begin{abstract}
 We develop the learning algorithm to build the architecture agnostic model of the reconfigurable optical interferometer. Programming the unitary transformation on the optical modes of the interferometer either follows the analytical expression yielding the unitary matrix given the set of phaseshifts or requires the optimization routine if the analytic decomposition does not exist. Our algorithm adopts the supervised learning strategy which matches the model of the interferometer to the training set populated by the samples produced by the device under study. The simple optimization routine uses the trained model to output the phaseshifts of the interferometer with the given architecture corresponding to the desired unitary transformation. Our result provides the recipe for efficient tuning of the interferometers even without rigorous analytical description which opens opportunity to explore new architectures of the interferometric circuits.
\end{abstract}

\maketitle

\section{Introduction}
Linear optical interferometers are rapidly becoming an indispensable tool in quantum optics \cite{Carolan2015} and optical information processing \cite{Harris2018}. The interest to linear optics grows due to broader availability of the integrated photonic fabrication technology to the scientific community. The key feature of the state-of-the-art integrated linear interferometer is the reconfigurability enabling the device to change its effect on the input optical mode upon the demand. This possibility has made the linear optical circuits particularly appealing for information processing challenges. In particular reconfigurable interferometers are the main ingredients of the contemporary linear optical quantum computing experiments \cite{Wang2018, Wang2019, Zhang2021} and are considered as the hardware accelerators for deep learning applications \cite{Hamerly2019, Wetzstein2020}. Furthermore, the fabrication quality and the improved scalability of the reconfigurable photonic circuits led to the emergence of the field-programmable photonic array concept - a multipurpose photonic circuit which can serve many possible applications by means of the low-level programming of the device \cite{PrezLpez2020}. 

The unitary transformation matrix $U$ completely describes the operation of the linear optical interferometer. The matrix $U$ couples the input optical modes of the device to the output ones $a^{(out)}_{j}=\sum_{j}U_{ij}a^{(in)}_{i}$. The architecture of the interferometer parametrizes the transformation $U = U(\{\varphi\})$ on the tunable parameters $\{\varphi\}$ which are typically the phase shifters controlling the relative phases between the arms of the interferometer. The architecture is labeled as universal if it allows reaching any arbitrary $N\times N$ unitary matrix by appropriately setting the phase shifts $\{\varphi\}$. The device programming is then essentially boiled down to establishing the correspondence between the desired matrix $U_{0}$ and the appropriate parameter set $\{\varphi^{0}\}$. Hurwitz analytical decomposition of the $N\times N$ unitary matrix \cite{Hurwitz1897} is the well-known example of the universal architecture. It implies straightforward implementation using simple optical components \cite{Reck1994, Clements2016} - the two-port Mach-Zender interferometers (MZI) with two controllable phase shifters. The interferometer architecture based on this decomposition is the mesh layout of the MZI which is very easy to program - the efficient inverse algorithm returns the values for each phase shifter given the unitary matrix $U_{0}$. The simplicity of this architecture comes at the cost of the extremely stringent fabrication tolerance. The universal operation is achieved if and only if the beamsplitters in the MZI blocks are perfectly balanced which is never the case in the real device. Numerical optimization methods have been adopted to mitigate the effect of the imperfections \cite{Burgwal2017, Dyakonov2018} but the simple programming flow is deprived.

The challenge to overcome the effect of the fabrication defects have also led to development of more sophisticated architectures \cite{Saygin2020, Fldzhyan2020} which have no simple analytical description and can only be programmed using the optimization routines. Running the optimization routine to set up the real physical device transformation requires the experimental execution of the resource-intensive procedure of the transformation matrix reconstruction\cite{Tillmann_2016} at each iteration of the optimization algorithm of choice. From the end user perspective the necessity to optimize the device each time when the transformation needs to be changed is unacceptable. Firstly, the optimization in the high-dimensional parameter space is itself a time-consuming procedure requiring sophisticated tuning and what's more there is no guarantee that the global minimum will be reached. Secondly, the algorithms providing fast convergence in multiparameter optimization problems are typically gradient-based and the precision of the gradient estimation of the objective function implemented by the physical device is limited by the measurement noise. Lastly, even though the number of switching cycles of the phase shifters is not strictly limited spending the device resource during tedious optimization procedures severely degrades the lifetime of the programmable circuit.

In this work, we develop the efficient algorithm for programming a linear optical inteferometer with complex architecture. We employ one of the main methods of machine learning - {\it supervised learning} of a numerical model, widely applied to the neural networks training \cite{Nielsen, Smooth2019, 1388456}. The model of the interferometer is learnt using the set of samples of transformations corresponding to different phase shifts. The trained model is used to quickly find the necessary phase shifts for a given unitary transformation using optimization routine applied to the model and not to the physical device. Our learning algorithm is divided into two stages: the training stage - find the model of the interferometer using the training set of sample transformations, and the programming stage - determine the phase shifts of the interferometer model corresponding to the required transformation.

\begin{figure*}[t!]
\center{\includegraphics[width=1.0\linewidth]{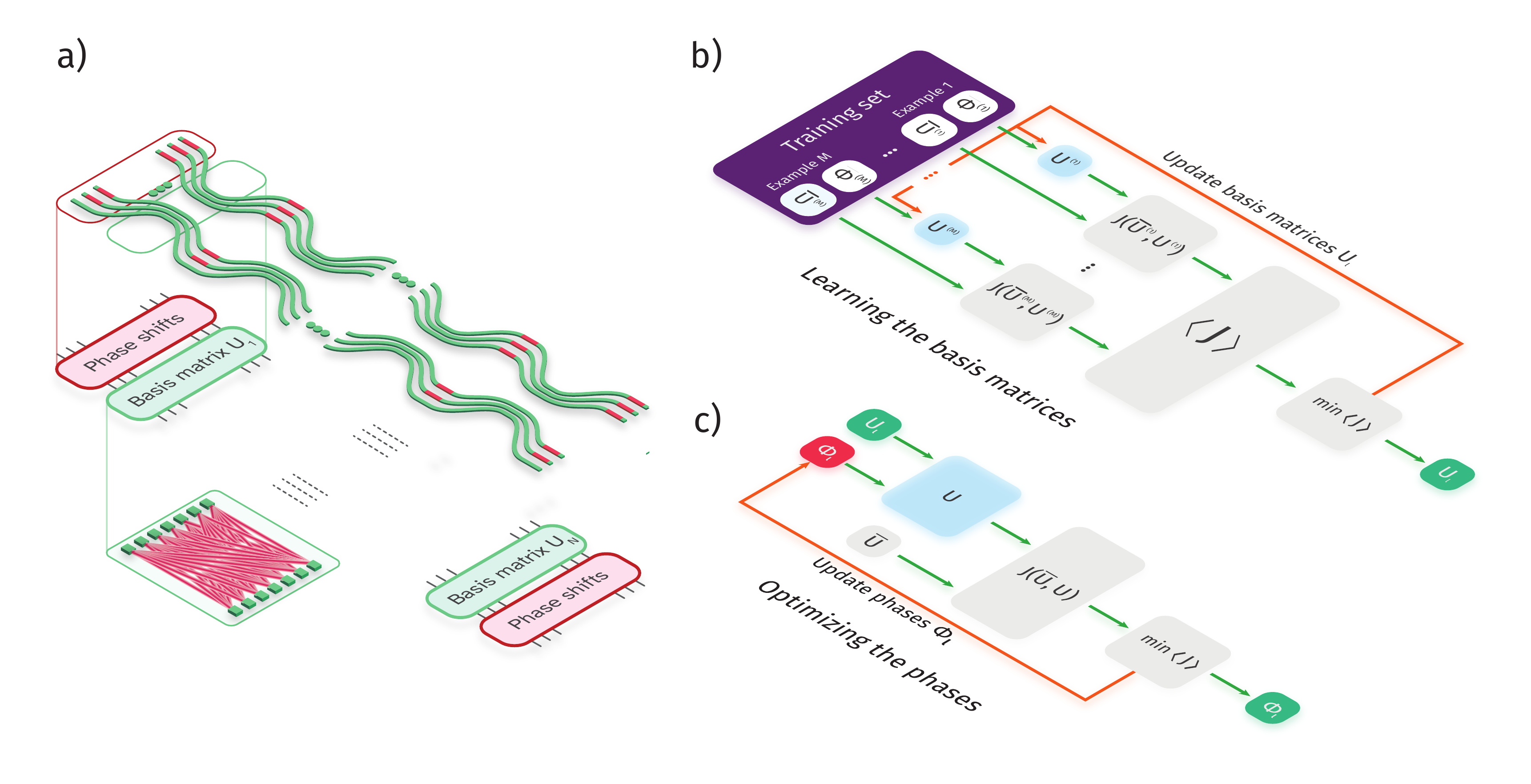}} \\
\caption{a) The schematic of the multimode interferometer structure and its integrated photonic implementation circuit. The basis matrices describe the distribution of light between the interferometer channels. The corresponding integrated elements may be implemented, for instance, as the waveguide lattices where the waveguides are coupled and thus the energy transfers between different waveguides \cite{Skryabin2021}. b) The workflow of the learning algorithm. The phase parameters $\bar{\Phi}^{(i)}$ from the training set are substituted to the expansion Eq. \ref{Expansion}. The output $U^{(i)}$ of the Eq. \ref{Expansion} and the $\bar{U}^{(i)}$ matrix from the training are used to calculate the figure of merit J. The optimization algorithm updates the guess for the basis matrices to minimize the distance averaged over the training set. c) The scheme of the interferometer tuning algorithm based on the learned model $\mathcal{M}$. The tuning process picks the phases $\Phi$ to minimize the distance $J$ between the required matrix $U_{0}$ and the model-implemented matrix $U(U_{\ell}^{\mathcal{M}},\Phi)$.}
\label{Interferometer}
\end{figure*}

\section{Formulation of the problem}

We devised our algorithm to solve the problem of programming the multimode interferometer consisting of alternating phaseshifting and mode mixing layers. This architecture has been proven to deliver close to universal performance and has no straightforward connection linking the elements of the matrix to the phase shifts of the interferometer \cite{Saygin2020}. This architecture serves as the perfect example to demonstrate the gist of our algorithm. The circuit topology is outlined in Fig. \ref{Interferometer}a). The unitary matrix $U$ is expressed as 
\begin{equation}\label{Expansion}
U = \Phi_{N + 1} U_{N} \Phi_{N} U_{N-1} \dots \Phi_{\ell + 1} U_{\ell} \Phi_{\ell} \dots \Phi_{2} U_{1} \Phi_{1},
\end{equation}
where $\Phi_{\ell} = \operatorname{diag} (e^{i \varphi_{{\ell}1}}, \dots, e^{i\varphi_{{\ell}{N}}})$, $\ell = 1, \dots, N+1$. We call $U_{\ell}$ the {\it basis} matrices because they completely define the interferometer operation. If the $U_{\ell}$ are available a simple numerical optimization routine finds the corresponding phase shifts $\varphi_{\ell k}$ thus completing the task to program the device. It is worth noting that the generalized form of the expansion of the type (\ref{Expansion}) given in \cite{Saygin2020} is valid for any linear-optical interferometer design. Indeed it is easy to confirm that every optical interferometer comprised of independent ingredients - fixed beamsplitting elements of any topology and phase modulators - can be unfolded into the sequence of unitary transformations coupling the modes of the circuit and the phase shifters. The only information required is the number of mode mixing layers and the phase shifting layers. The fact that the inner structure of the beamsplitting elements can be arbitrary and there is no restriction on the length, order or number of mode mixers and phase shifters gives us the strong ground to call our algorithm architecture agnostic. 

The problem underpinning the difficulty of programming the required unitary in the multimode architecture is that the basis matrices $U_{\ell}$ of the fabricated device do not match the ones implied by the circuit optical design. The interferometer universality is not degraded but efficient evaluation of the phase shifts $\varphi_{\ell k}$ becomes impossible since $U_{\ell}$ are not know anymore. This brings us to the first step of our learning algorithm - the reconstruction of the parameters of the basis matrices utilizing the information from the training set $\mathcal{T}$ gathered from the device of interest (see Fi.\ref{Interferometer}b)). The set $\mathcal{T}$ includes $M$ pairs $(\bar{U}^{(i)},\bar{\Phi}^{(i)})$ obtained by seeding the device with random phase shifts $\bar{\Phi}^{(i)}$ and applying the unitary reconstruction procedure of choice \cite{Tillmann_2016, Suess2020} to get the $\bar{U}^{(i)}$. The basis matrices $U_{\ell}$ are then determined as the solution of the optimization problem
\begin{align}\label{eq:training_optimization}
    \{U_{\ell}\} = \underset{\{U_{\ell}\}}{argmin} \langle J(\bar{U},U(\{U_{\ell}\},\bar{\Phi}))\rangle_{\mathcal{T}},
\end{align}
where the figure of merit is averaged over the training set $\mathcal{T}$. Once the basis matrices are determined we move to the second step of the algorithm (see Fig.\ref{Interferometer}c)) - finding the appropriate phase shifts $\varphi_{{\ell k}}$ which will adjust the interferometer model $\mathcal{M}$ to match the unitary matrix $U_{0} \notin \mathcal{T}$.

% \begin{figure}[t]
% \center{\includegraphics[width=0.8\linewidth]{Tuning/tuning_scheme.pdf}} \\
% \caption{The scheme of the interferometer tuning algorithm based on the learned model $\mathcal{M}$. The tuning process picks the phases $\Phi$ to minimize the distance $J$ between the required matrix $U_{0}$ and the model-implemented matrix $U(U_{\ell}^{\mathcal{M}},\Phi)$.}
% \label{Tuning_task}
% \end{figure}

\section{The learning algorithm}
In this section, we present the algorithm which learns the interferometer model $\mathcal{M}$ based on the initial data contained in the training set $\mathcal{T}$. We reduce the learning problem to the multiparameter optimization of the nonlinear functional $J$. In this section we present the mathematical framework of the learning algorithm and exemplify its performance on the multimode interferometer.

\subsection{The figure of merit}
The figure of merit $J$ to be studied in our work is the Frobenius norm $J_{FR}$: 
\begin{equation}
J_{FR} (U, \bar{U}) \equiv \dfrac{1}{N} \sum_{i, j = 1}^N |u_{ij} - \bar{u}_{ij}| ^2.
\label{eq:frobenius_original}
\end{equation}
It is invariant only under the identity transformation, that is, $J_{FR} (U, \bar{U}) = 0$ only if the the magnitude and the phase of $U$ and $\bar{U}$ matrix elements are identical. The expression \ref{eq:frobenius_original} can be rewritten using Hadamard's product $(A \odot B)_{i, j}=(A)_{i, j} \cdot(B)_{i, j}$ operation and takes the following form:
\begin{equation}
\hspace{0.5cm} J_{FR} (U, \bar{U}) = \dfrac{1}{N}\sum_{i, j = 1}^N ((U - \bar{U}) \odot (U - \bar{U})^{*})_{i,j}. \hspace{0.5cm}
\label{eq:frobenius_hadamard_form}
\end{equation}

The gradient of the $J_{FR}$ with respect to the parameter set
$\{\bar{\alpha}\}$ is given by:

\begin{equation}
\partial_{\bar{\alpha}} J_{FR}(U(\bar{\alpha}), \bar{U}) = \dfrac{2}{N} Re \sum_{i, j = 1}^N ((U - \bar{U})^{*} \odot \partial_{\bar{\alpha}} U)_{i, j}.
\label{eq:Grad_FR}
\end{equation}

\subsection{Computing the gradients of $J$}

The gradient-based optimization algorithm substantially benefit from the analytical gradient expressions of the optimized functions. It turns out that the multimode interferometer expansion \ref{Expansion} admits simple analytic forms of the gradients over the $u_{ij}$ elements of the basis matrices $U_{\ell}$ and over the phase shifts $\varphi_{\ell k}.$ We will derive the analytical expressions of the gradients $\partial_{x^{(\ell)}_{ij}} J_{FR}$, $\partial_{y^{(\ell)}_{ij}} J_{FR}$ and $\partial_{\varphi_{\ell k}} J_{FR}$ required during learning and tuning stages of the algorithm respectively. The Eq. \ref{eq:Grad_FR} stems that the gradients $\partial_{x^{(\ell)}_{ij}} J_{FR}$, $\partial_{y^{(\ell)}_{ij}} J_{FR}$ and $\partial_{\varphi_{\ell k}} J_{FR}$ calculation is reduced to finding the expressions for $\partial_{x^{(\ell)}_{ij}} U$, $\partial_{y^{(\ell)}_{ij}} U$ and $\partial_{\varphi_{\ell k}} U$ respectively. 

We will first focus on the $\partial_{x^{(\ell)}_{ij}} J_{FR}$ and $\partial_{y^{(\ell)}_{ij}} J_{FR}$ gradients. In order to simplify the computation we introduce $N$ auxiliary matrices $A_{\ell}$ and another $N$ auxiliary matrices $B_{\ell}$ as the partial products taken from the expansion Eq. \ref{Expansion}: 
\begin{equation}\label{eq:aux_matrix_definition}
U = A_{\ell} U_{\ell} B_{\ell}.    
\end{equation}
where $A_{\ell}$ and $B_{\ell}$ can be calculated iteratively:
\begin{equation}\label{eq:Aux_matrix1}
\begin{cases}
   A_N = \Phi_{N+1}, 
   \\
   A_{\ell} = A_{{\ell} + 1}(U_{{\ell} + 1}\Phi_{{\ell}+1}), \hspace{0.5cm} {\ell} = N - 1, \ldots ,1,
   \\
   B_1 = \Phi_1, 
   \\
   B_{\ell} = (\Phi_{{\ell}} U_{{\ell} - 1})B_{{\ell} - 1}, \hspace{0.8cm} {\ell} = 2, \ldots ,N.
\end{cases}
\end{equation}

Next, given that $x_{ij}^{({\ell})}$ and $y_{ij}^{({\ell})}$ are the real and imaginary parts of $u_{ij}^{({\ell})}$ respectively we get the expressions for the gradients:
\begin{equation}\label{Training_task}
\hspace{0.3cm}\dfrac{\partial U}{\partial x_{ij}^{({\ell})}} = A_{\ell} \Delta^{(ij)} B_{\ell} \hspace{0.2cm} \text{ and }
\hspace{0.3cm}\dfrac{\partial U}{\partial y_{ij}^{({\ell})}} = i A_{\ell} \Delta^{(ij)} B_{\ell},
\end{equation}
where $\Delta^{(ij)}$ are the matrices in which all elements are zeros, except $\Delta^{(ij)}_{ij}=1$. The Appendix \ref{app:gradients} provides the detailed derivation of the Eq. \ref{Training_task}.

Once the basis matrices of the model $M$ are learnt we can use them to calculate the gradients $\partial_{\varphi_{\ell k}} J_{FR}$. The derivation of the $\partial_{\varphi_{\ell k}} J_{FR}$ also requires to introduce The $N + 1$ auxiliary matrices $C_{\ell}$ and $N + 1$ matrices $D_{\ell}$
\begin{equation}
    U = C_{\ell} \Phi_{\ell} D_{\ell}
\end{equation}
representing the partial products from the general expansion Eq. \ref{Expansion}. The iterative formula establishes $C_{\ell}$ and $D_{\ell}$ for each index $\ell$:
\begin{equation}\label{eq:Aux_matrix2}
 \begin{cases}
   C_{N+1} = I, 
   \\
   C_{\ell} = C_{{\ell} + 1}(\Phi_{{\ell} + 1}U_{\ell}), \hspace{0.5cm} {\ell} = N, \ldots ,1,
   \\
   D_1 = I, 
   \\
   D_{\ell} = (U_{{\ell} - 1} \Phi_{{\ell} - 1})D_{{\ell} - 1}, \hspace{0.8cm} {\ell} = 2, \ldots , N+1.
 \end{cases}
\end{equation}
Once the $C_{\ell}$ and $D_{\ell}$ are computed, the gradients  $\partial_{\varphi_{\ell k}} U$ are given by
\begin{equation}\label{Tuning_grad}
\hspace{0.3cm}\dfrac{\partial U}{\partial \varphi_{\ell k}} = i e^{i \varphi_{\ell k}} C_\ell \Delta^{(kk)} D_{\ell}, \hspace{0.3cm}
\end{equation}
where all $\Delta^{(kk)}$ elements are zeros, except the $\Delta^{(kk)}_{kk}=1$.
The details of the multimode interferometer architecture tuning is described in \cite{Saygin2020}. The output of the $\varphi_{\ell k}$ consludes the workflow of the algorithm.

\section{Numerical experiment}

In this section we provide key performance metrics of the interferometer model learning algorithm. We test the algorithm scaling properties with respect to the training set $M$ size and the number of interferometer modes $N$. To certify the quality of the model $\mathcal{M}$ we employ the cross-validation methodology - the quality tests use another set of examples which wasn't included in $\mathcal{T}$.

The simulation of the learning algorithm follows a series of steps. We first generate the training set $(\bar{U}^{(i)},\bar{\Phi}^{(i)})$ using the multimode interferometer expansion Eq. \ref{Expansion}, while we choose the phases randomly from a uniform distribution from $0$ to $2\pi$. The basis matrices $U_{\ell}$ are sampled randomly from the Haar-uniform distribution using the QR decomposition \cite{mezzadri2006generate}. In the real-life setting the elements of $\mathcal{T}$ are the outcomes $\bar{U}^{(i)}_{rec}$ of the unitary reconstruction \cite{Tillmann_2016, Suess2020} algorithms applied to the reconfigurable interferometer programmed with the phases $\bar{\Phi}^{(i)}$. The subtleties of gathering the appropriate training set experimentally are discussed in Sec. \ref{sec:discussion}. 

The proper interferometer model must accurately predict the unitary matrix of the real device with the certain set of phases $\Phi$ applied. The cross-validation purpose is to estimate the predictive strength of the model $M$. For cross-validation we generate the test set $(\hat{U}^{(i)},\hat{\Phi}^{(i)})$ comprised of randomly selected phases $\hat{\Phi}^{(i)}$ and the corresponding $\hat{U}^{(i)}$. The cross-validation test uses each sample from the test set to verify whether the interferometer model with phases $\hat{\Phi}^{(i)}$ outputs the unitary $\hat{U}^{(i)}$. The model $U$ is considered to pass the cross-validation if $J(U,\hat{U}) \leq 10^{(-2)}$. The criteria has been derived empirically by analyzing the behaviour of $J_{FR}$ convergence on the test set. If the model passes cross-validation the $J_{FR}(U,\hat{U})$ experiences rapid decrease down to the values less than $10^{-2}$.

The model is initialized with basis matrices selected either randomly or with \textit{a priori} knowledge available based on the design of the physical elements realizing the basis matrices. We will study both cases and refer to the random initialization as the \textit{black box} model. At each epoch of the learning process we estimate the average gradient over the collection of examples from $\mathcal{T}$ and update the basis matrices according to the optimization algorithm (stochastic L-BFGS-B algorithm, SciPy package). Instead of averaging the gradient over full training set we randomly draw a subset of $m=5$ pairs $(\bar{U}, \bar{\Phi})$ each epoch and use this subset for averaging. The value $m=5$ has been determined empirically as it was providing substantial computational speed-up while still keeping high training accuracy. We don't use the unitary parametrization of the basis matrices during the learning procedure and these matrices simply as the complex-valued square matrices. Since the parameters of the model are then the real $x_{ij}^{(\ell)}$ and the imaginary $y_{ij}^{(\ell)}$ parts of each basis matrix $U_{\ell}$ the updated basis matrices don't belong to the unitary space. We use the polar decomposition $A = H V$, where $H$ is a hermitian and $V$ is the unitary matrix, to project updated complex basis matrix $C_{\ell}$ onto the closest unitary $U_{\ell}$ at each step of the optimization \cite{Fan1955}. This method helps to avoid local minimum problem which may arise due to sophisticated unitary matrix parametrization.

The simulation code is written in Python employing Numpy and Scipy packages. The code is publicly available on GitLab \cite{Kuzmin2020}.

\subsection{Model - black box}\label{Black_box}

\begin{figure}[t!]
\begin{minipage}[h]{0.49\linewidth}
\center{\includegraphics[width=1.11\linewidth]{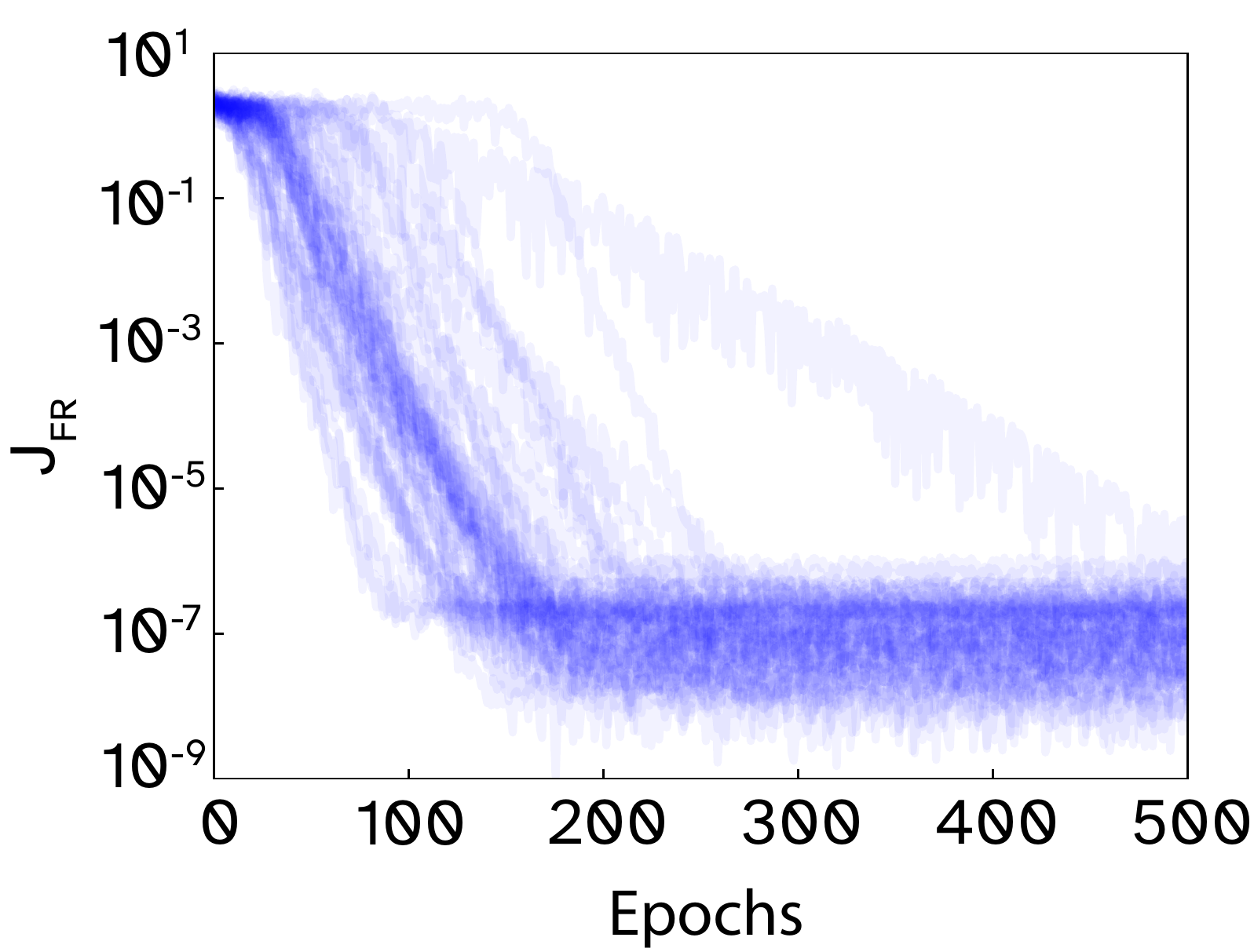}} a) $N=4$.\\
\end{minipage}
\hfill
\begin{minipage}[h]{0.49\linewidth}
\center{\includegraphics[width=1.11\linewidth]{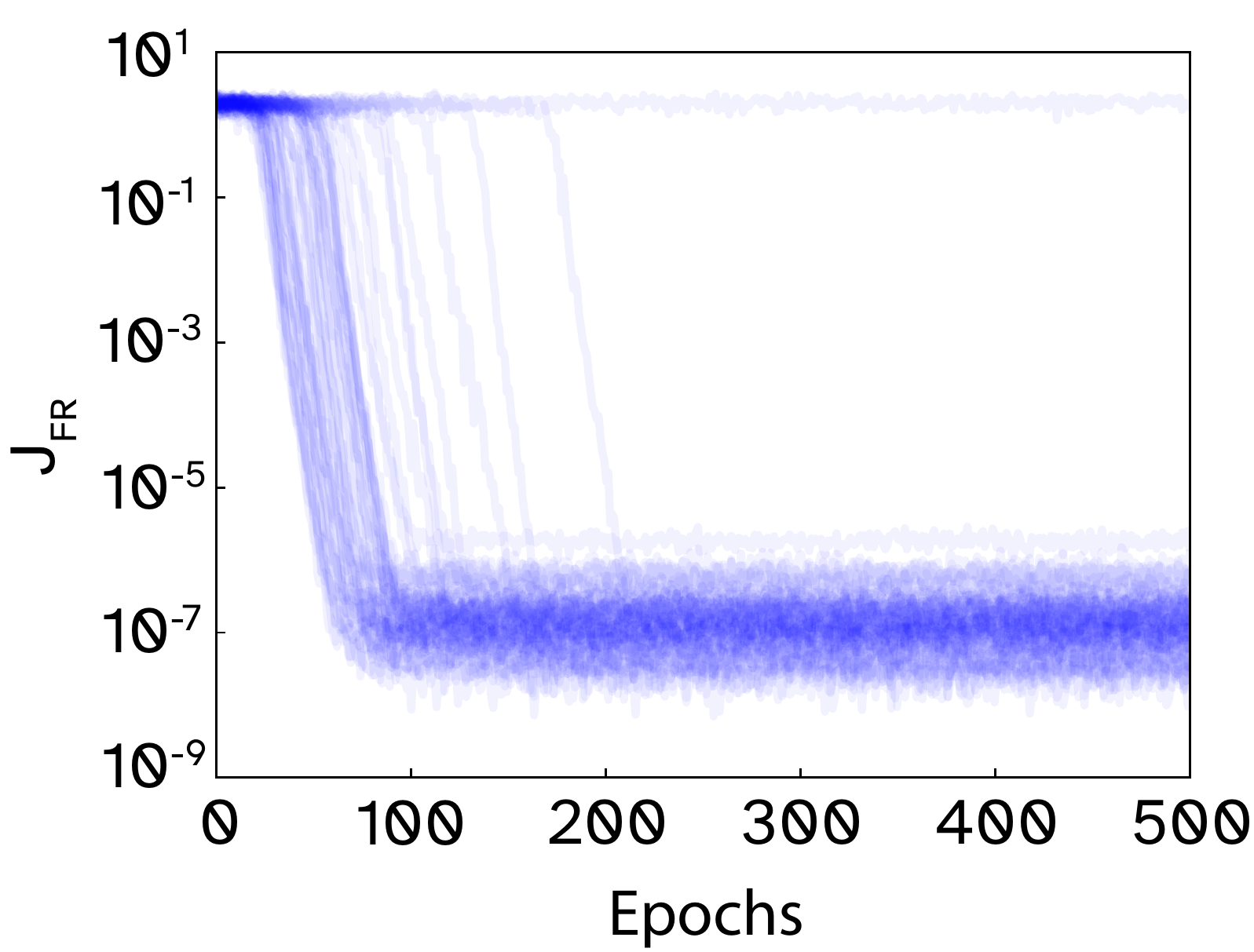}} b)$N=5$.\\
\end{minipage}
\vfill
\begin{minipage}[h]{0.49\linewidth}
\center{\includegraphics[width=1.11\linewidth]{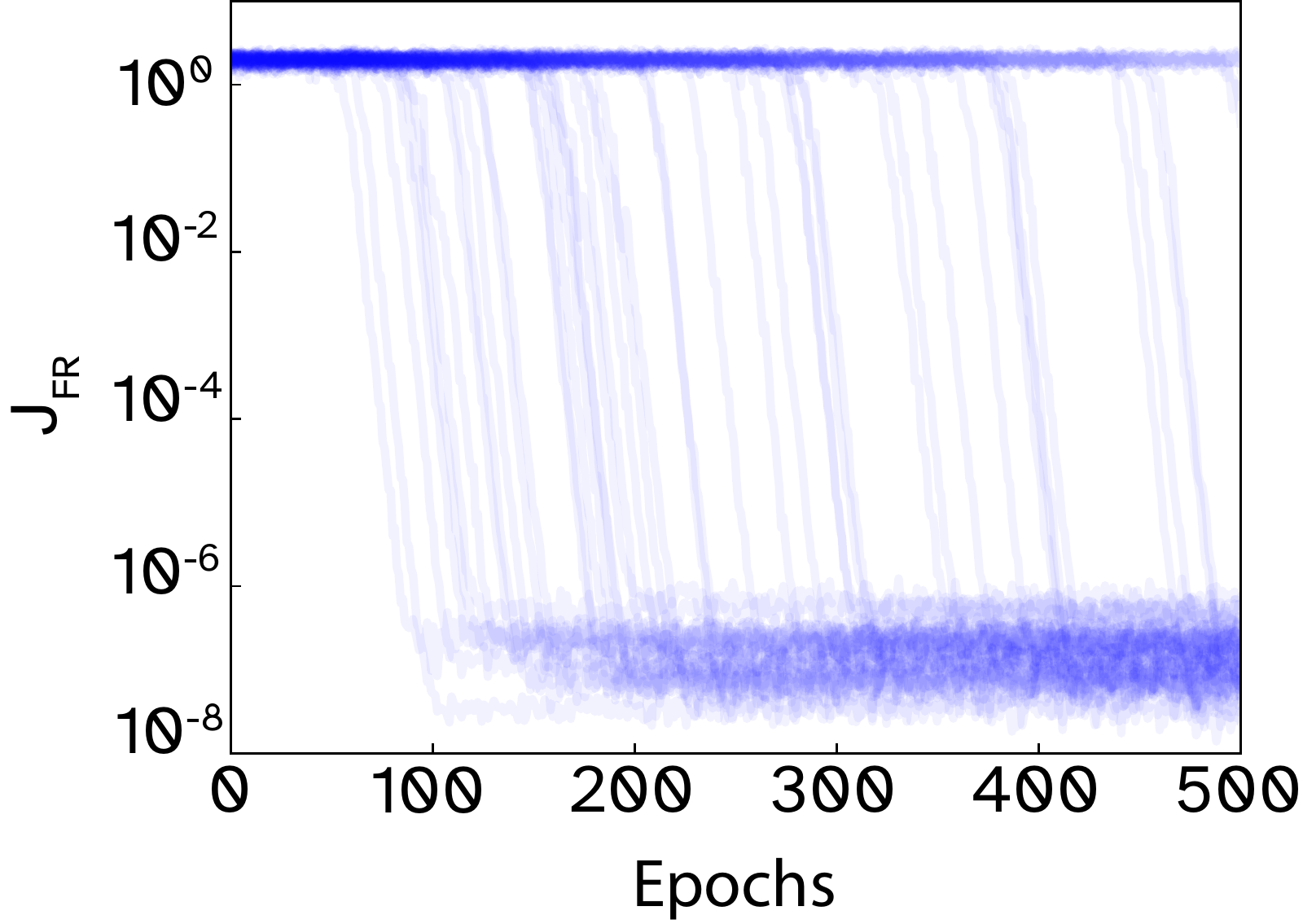}} c)$N=6$.\\
\end{minipage}
\begin{minipage}[h]{0.49\linewidth}
\center{\includegraphics[width=1.11\linewidth]{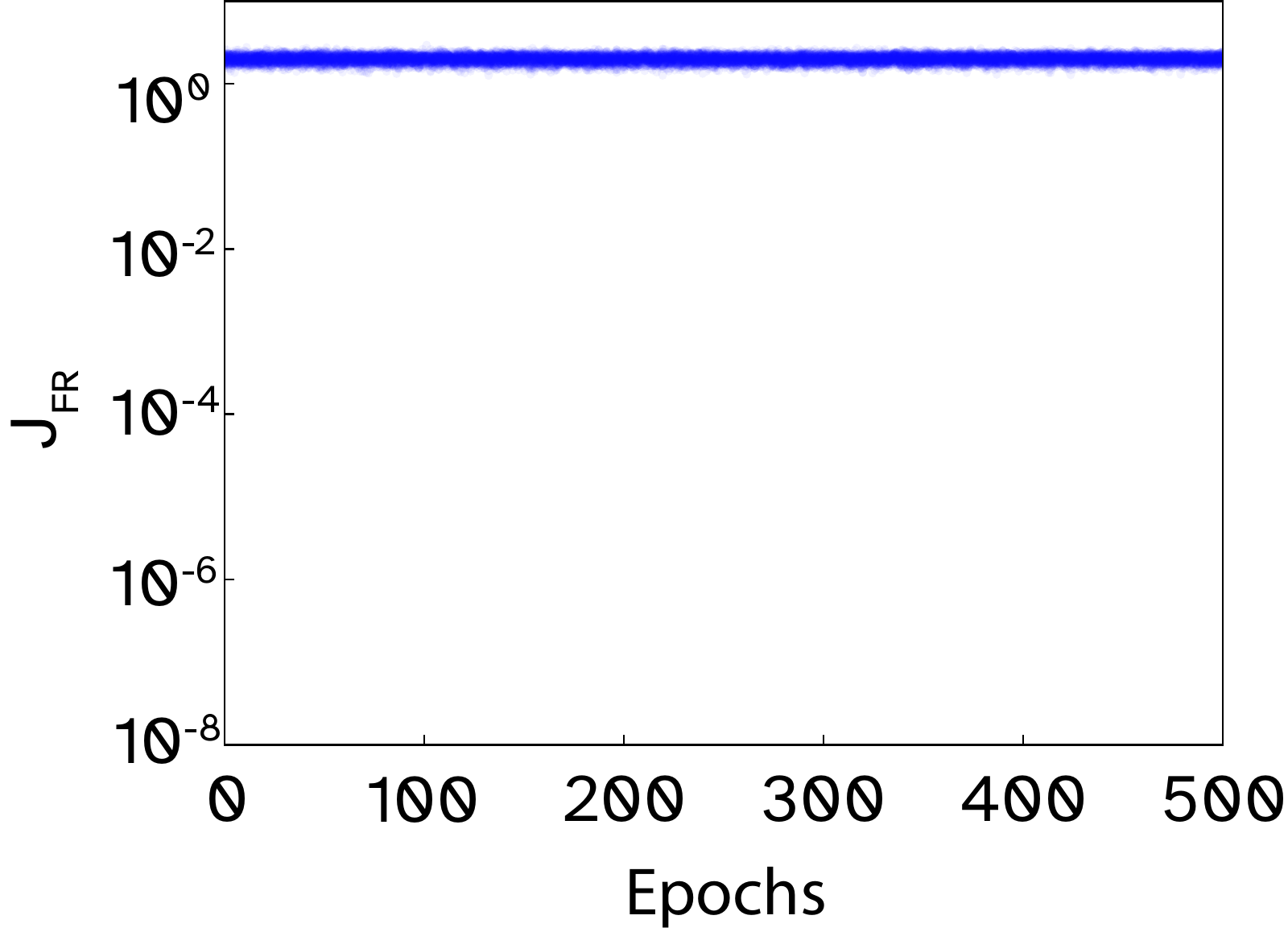}} d)$N=7$.\\
\end{minipage}
\caption{The learning examples for different numbers of optical modes $N$. We plot the Frobenius functional convergence on the test dataset on the epoch number for $50$ different instances of the optimization algorithm (for each instance we randomly chose the initial basis matrices $U_{\ell}$). Case a) - the number of optical modes $N = 4$, the size of the training dataset $M = 7$, b) - $N = 5$, $M = 30$, c) - $N = 6$, $M = 230$, d) - $N = 7$, $M = 2000$.}
\label{Training_examples}
\end{figure}

\begin{figure}[t!]
\begin{minipage}[h]{0.32\linewidth}
\center{\includegraphics[width=1\linewidth]{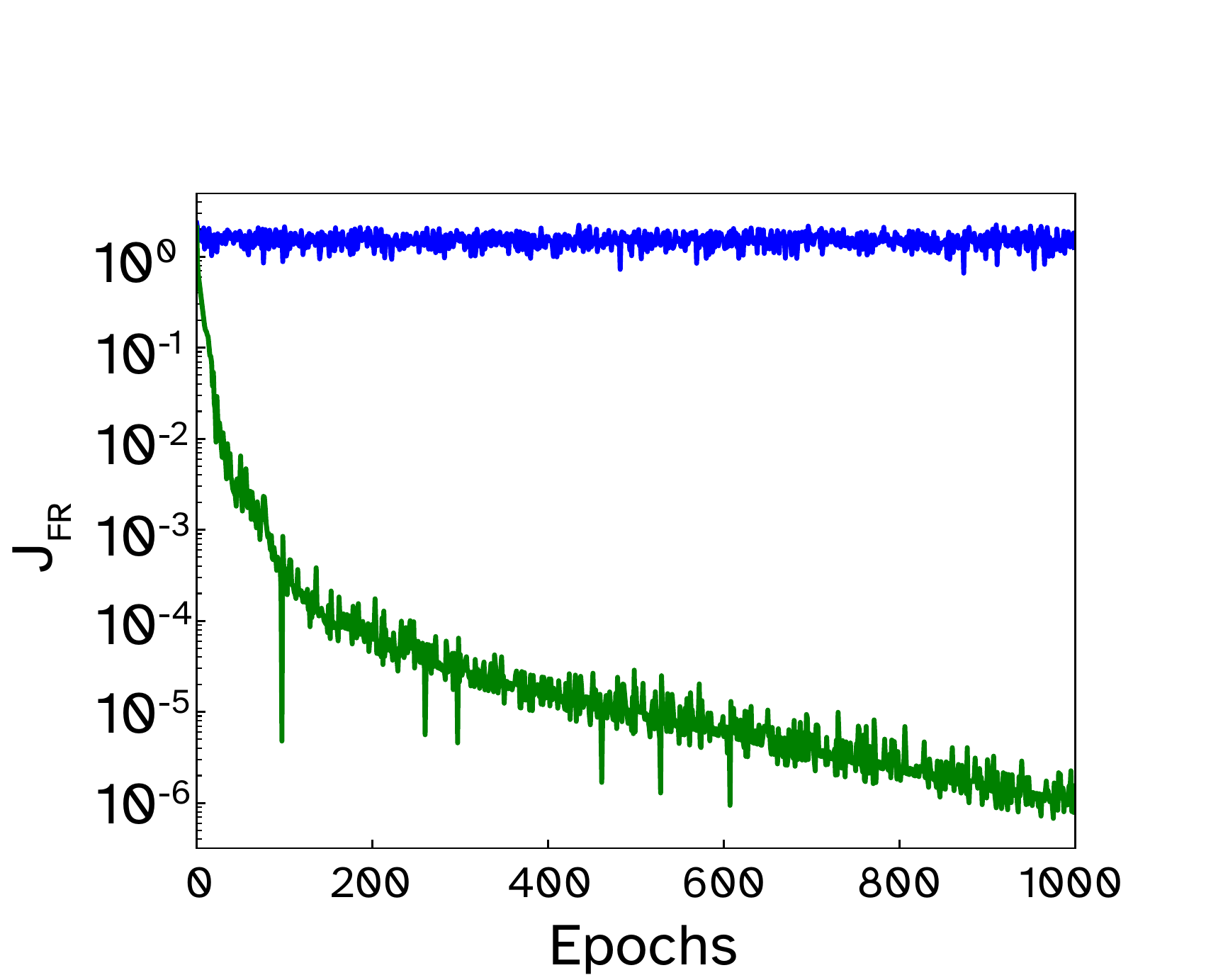}} a) $M = 2$\\
\end{minipage}
\hfill
\begin{minipage}[h]{0.32\linewidth}
\center{\includegraphics[width=1\linewidth]{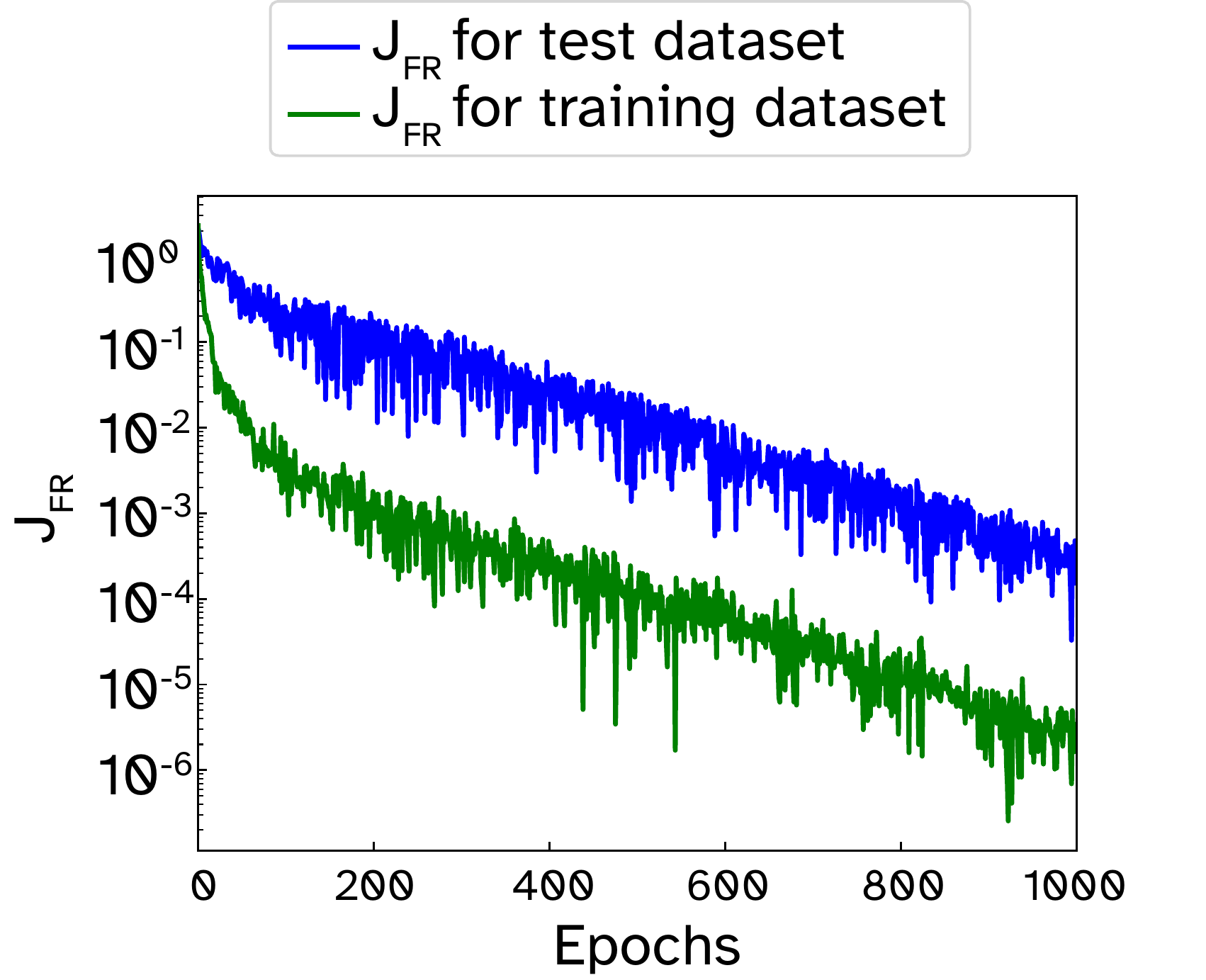}} b) $M = 3$\\
\end{minipage}
\hfill
\begin{minipage}[h]{0.32\linewidth}
\center{\includegraphics[width=1\linewidth]{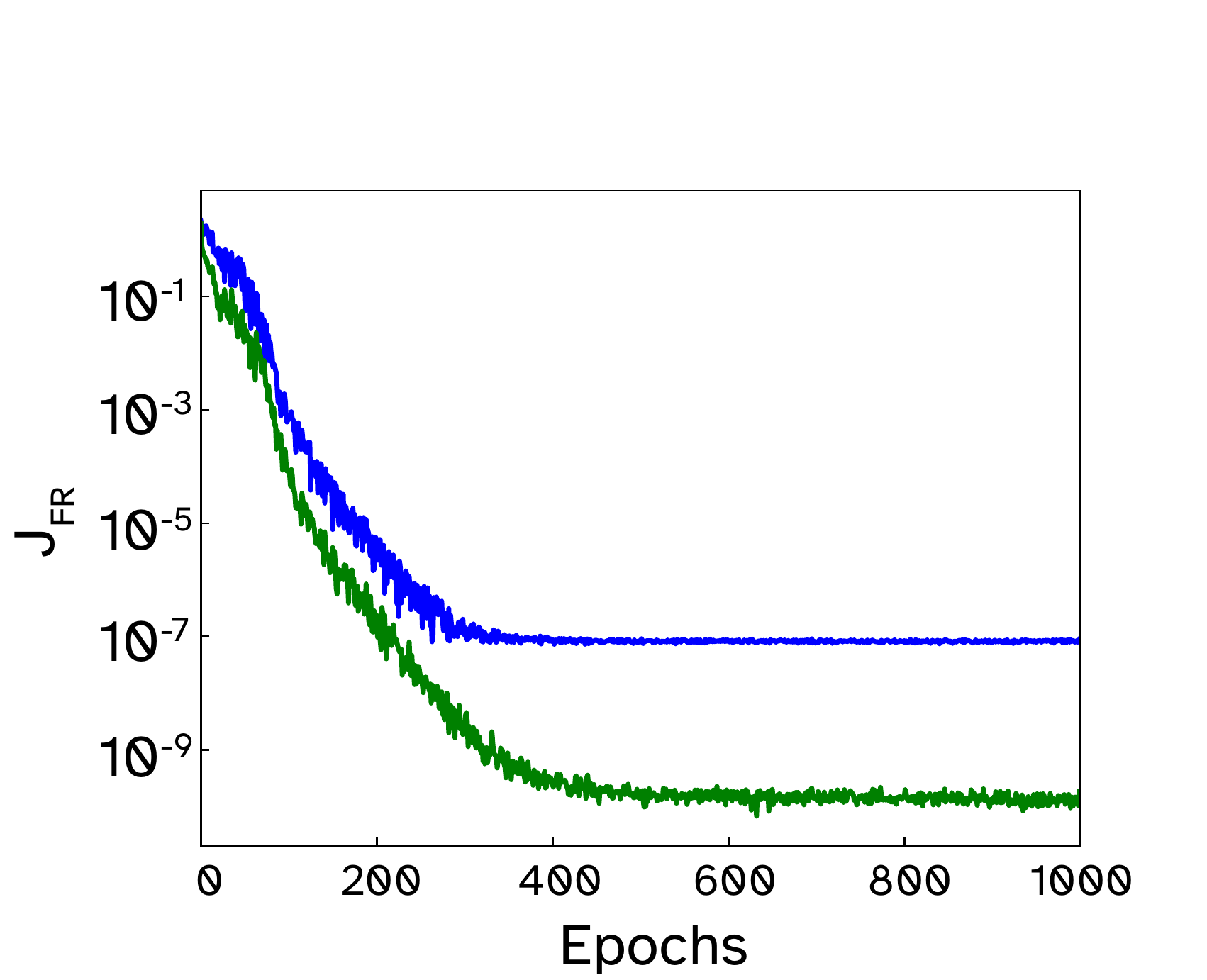}} c) $M = 4$\\
\end{minipage}
\begin{minipage}[h]{0.45\linewidth}
\center{\includegraphics[width=1\linewidth]{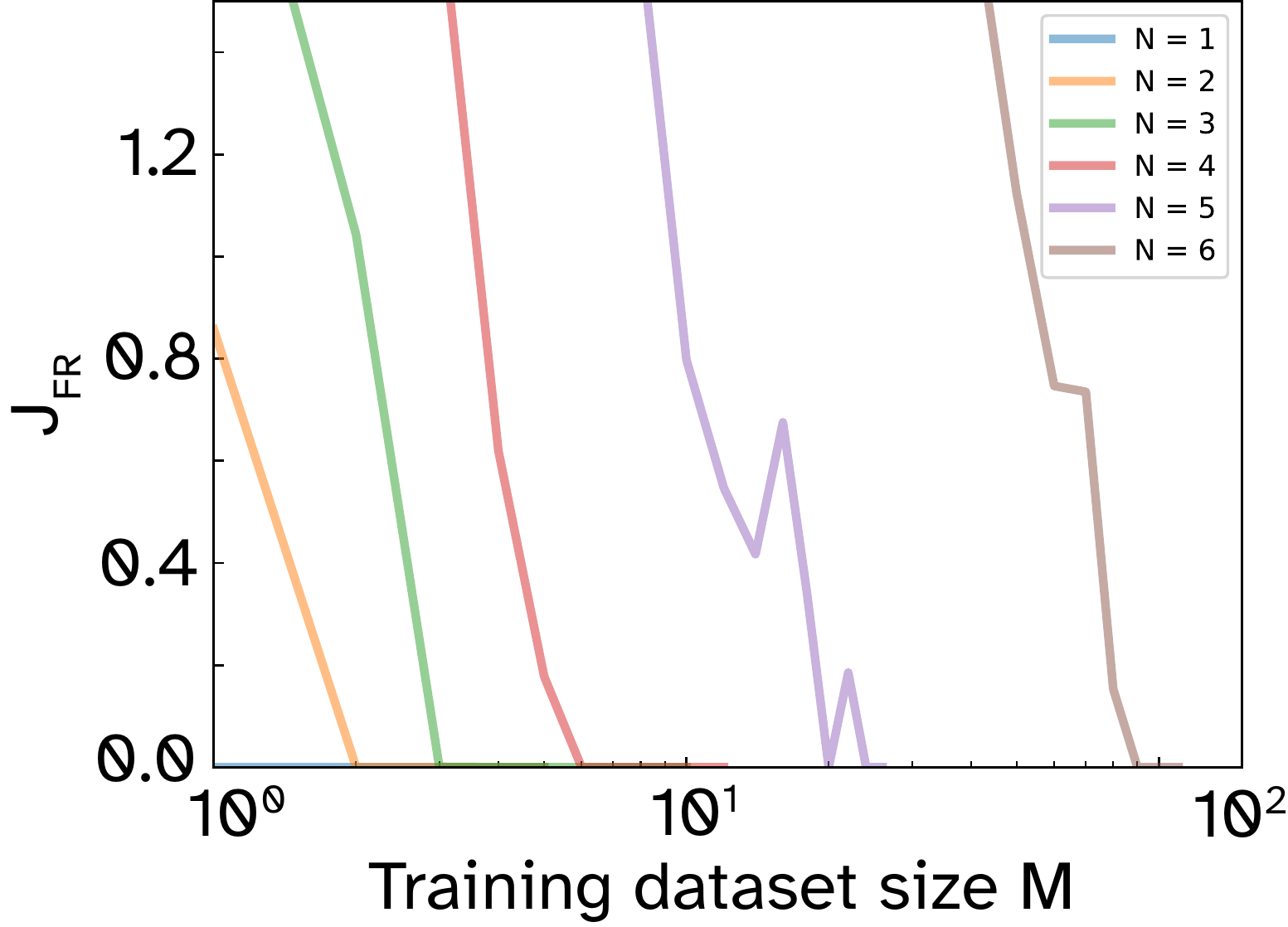}} d) \\
\end{minipage}
\hfill
\begin{minipage}[h]{0.45\linewidth}
\center{\includegraphics[width=1\linewidth]{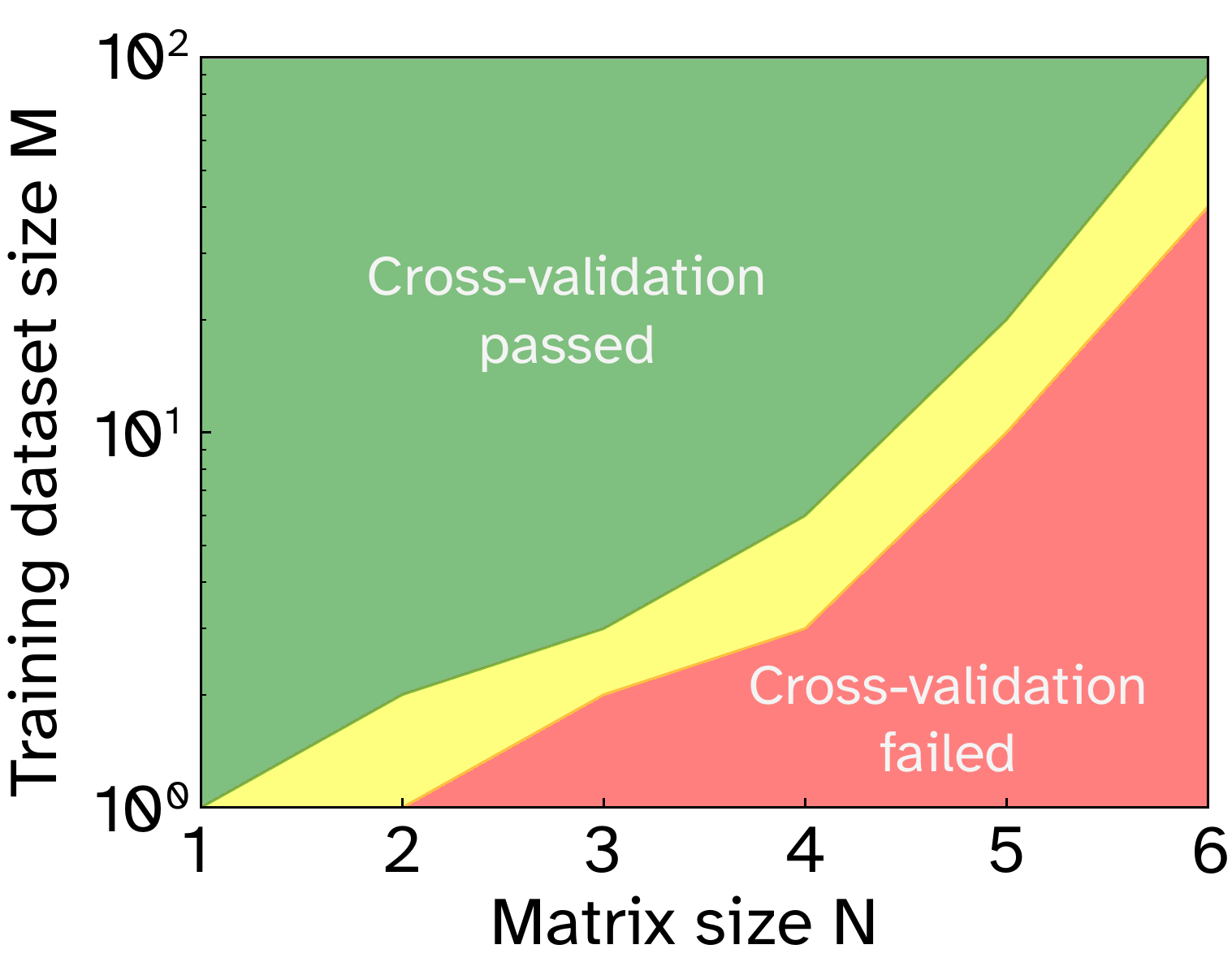}} e) \\
\end{minipage}
\caption{Top row illustrates the cross-validation convergence of the $N = 3$ interferometer during the learning based on the set $\mathcal{T}$ with different sizes. Case a) - the model does not pass cross-validation, b) - intermediate case, c) - the model passes the cross-validation test. The panel d) in the bottom row shows the dependence of the optimized $J_{FR}$ averaged over 10 instances for different numbers of modes $N$ and different sizes of the training dataset $M$. The picture in the panel e) maps the region where the model passes the cross-validation criteria.}
\label{fig:training_set}
\end{figure}

We start first from considering the \textit{black box} model. This scenario implies no {\it a priori} information about the basis matrices $U_{\ell}$ is available and the interferometer is represented by a black box type system with $N^2$ variable phase parameters $\varphi_{{\ell}k}$. Therefore, the model should be initialized with completely random guess. The initial basis matrices are sampled from the Haar-random unitary distribution \cite{mezzadri2006generate}. Fig. \ref{Training_examples} illustrates the convergence of the average value of the functional $J_{FR}$ during the learning for different values of interferometer dimension $N$. The model cross-validation testing (see Fig. \ref{fig:training_set}) determines the size $M$ of the training set for each value of $N$.

It should be noted that as $N$ increases a {\it plateau} appears, which heavily impacts the learning convergence. The plateau becomes significant already at $N = 6$ (Fig. \ref{Training_examples}c). For $N > 6$, we failed to observe learning in the black box - the average value of the figure of merit remains at a plateau all the time. The work \cite{pascanu2014saddle} suggests that the reason for the plateau in high-dimensional optimization problems is the presence of the large number of saddle points in the optimized function landscape rather than local minima. Several algorithms exploiting the idea of the adaptive gradient have been developed to tackle the problem of escaping the plateau \cite{staib2020escaping, Ba2014}.

Until this moment the elements of $\mathcal{T}$ included the matrices $\bar{U}^{(i)}$ which were artificially generated using the Eq. \ref{Expansion} initialized with the $\bar{\Phi}^{(i)}$ set of phases. Gathering the same set using the real device means that the reconstruction of the $\bar{U}^{(i)}_{exp}$ matrices must be performed with absolute precision, which is never the case in the experiment. The learning algorithm has to be designed to tolerate the certain amount of discrepancy between the ideal $\bar{U}_{ideal}$ and the reconstructed $\bar{U}_{exp}$ matrices. These deviations are the inevitable consequence of imperfections of measurement tools used during the reconstruction process. We have modeled the behaviour of the learning algorithm seeded with a training set including the phase shifts $\bar{\Phi}^{(i)}$ and the unitaries $\bar{U}^{(i)}_{exp}$ slightly deviated from their theoretical counterpart $\bar{U}^{(i)}_{ideal}$.

\begin{figure}[t!]
\begin{minipage}[h]{0.49\linewidth}
\center{\includegraphics[width=1.11\linewidth]{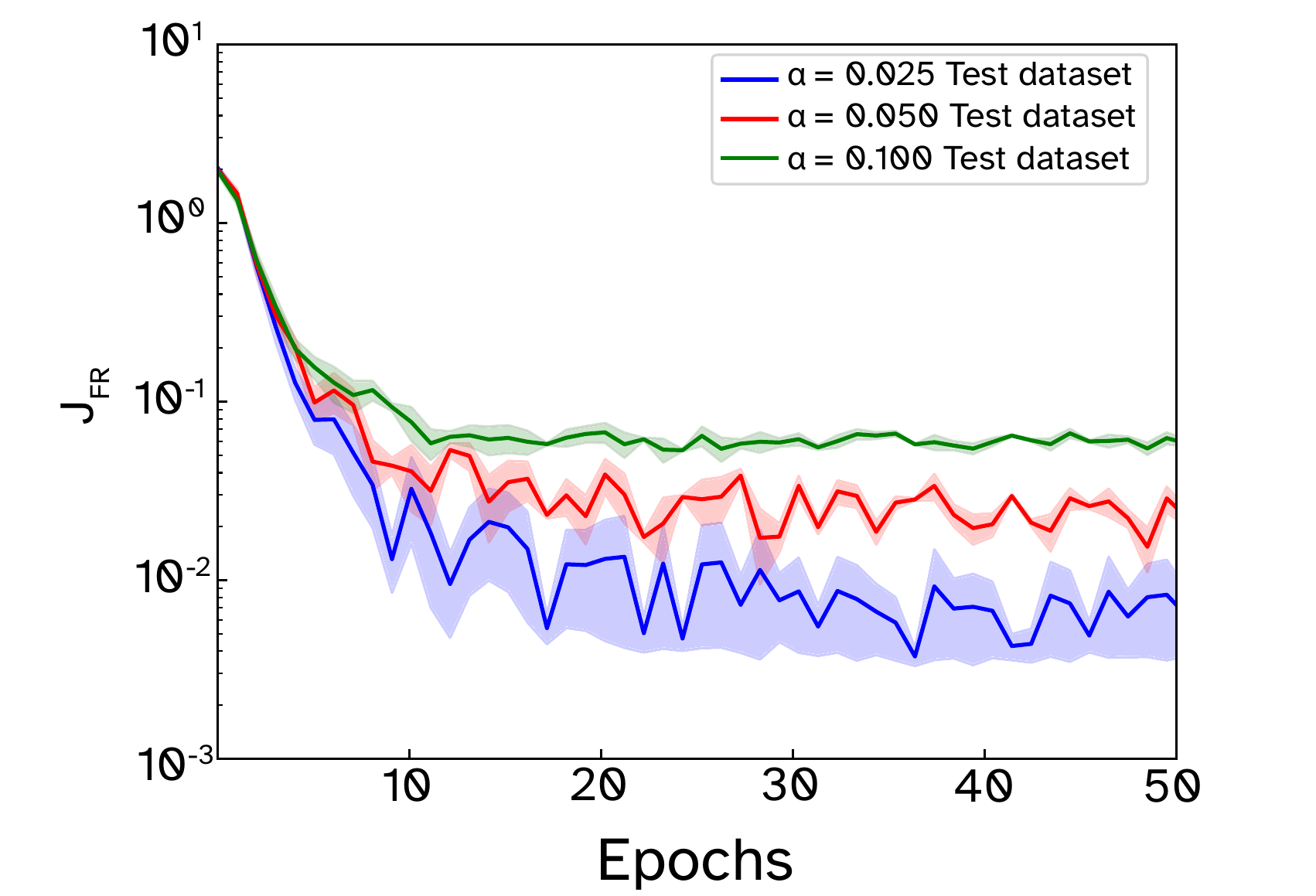}} a) \\
\end{minipage}
\hfill
\begin{minipage}[h]{0.49\linewidth}
\center{\includegraphics[width=1.11\linewidth]{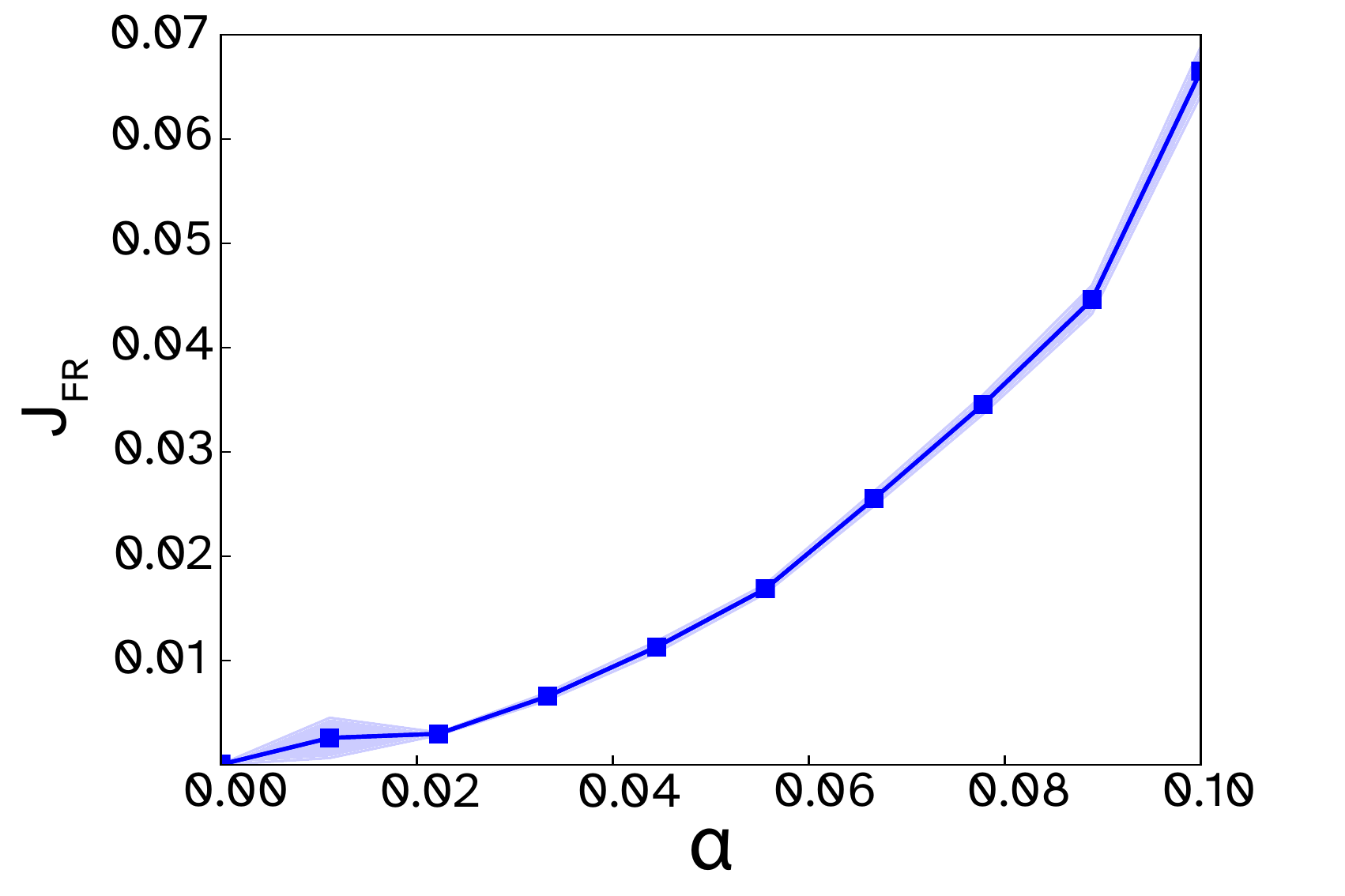}} b) \\
\end{minipage}
\caption{The effect of the imperfect training set on the interferometer model. a) - The convergence plots for the $N = 2$ interferometer models supplied with $M = 5$ training set, sampled with $\alpha = 0.025,\,0.05,\,0.1$. b) - The average value of the Frobenius functional $J_{FR}$ computed in the cross-validation test. Averaging was performed with $1000$ models learnt using the training sets corresponding to the different basis matrices $U_{\ell}^{(0)}$. For each model we performed the cross-validation test with $1000$ phase shift sets.}
\label{Noisy_analytics}
\end{figure}

The deviation is introduced between $\bar{U}_{exp}$ and $\bar{U}_{ideal}$ as the polar projection \cite{Fan1955} of the perturbed $\bar{U}_{ideal}$ onto the unitary space:

\begin{equation}\label{eq:deviation}
    A = \bar{U}_{ideal} + \alpha(X + i Y), \hspace{0.3cm} A = H\bar{U}_{exp},
\end{equation}

where $X$ and $Y$ are the random real-valued matrices of size $N \times N$, which elements are sampled from the normal distribution $\mathcal{N}(0,\,1)$. The degree of deviation is controlled by the real-valued parameter $\alpha$. The calibration curves provided in the Appendix \ref{app:deviation} juxtapose the more common matrix distance measure the fidelity $F$ calculated as the Hilbert-Schmidt scalar product  to the deviation values $F(\alpha)$. These curves should develop the intuition to interpret the $J_{FR}$ values. The Fig. \ref{Noisy_analytics} illustrates the the convergence of the model of the simplest case $N=2$ supplied with the training set sampled with the given deviation $\alpha$. The $\alpha \approx 0.04$ indicates the threshold at which the model fails to pass the cross-validation criteria $J_{FR} \leq 10^{-2}$. 

\subsection{The model with \textit{a priori} knowledge}

The \textit{black box} model results expounded in the sec. \ref{Black_box} evidence that the optimization complexity of the model with arbitrary initial basis matrices grows rapidly in the training set volume $M$. In this section we study the choice of the initial approximation for the basis matrices $U_{\ell}$ which enables learning for the larger dimension $N$. The case when the basis matrices $U_{\ell}$ are completely unknown does not adequately reflect the real situation. In practice the optical circuits with well-defined geometry and optical properties implements the basis matrices. The prototypes of these circuit can be tested beforehand to verify the performance of the circuit including the mode transformation that it generates. Contemporary integrated photonics fabrication technologies guarantee the reproducibility up to the certain level of precision specific to each technological line. Summing up the basis matrix unitary transformation $U_{\ell}^{est}$ can be estimated in advance. This estimate serves as the initial guess for the optimization algorithm at the model training stage. In this section we will demonstrate how this knowledge substantially simplifies the optimization and enables learning the models of the interferometers with $N$ up to at least few tens of modes.

\begin{figure}[t!]
\begin{minipage}[h]{0.48\linewidth}
\center{\includegraphics[width=1.11\linewidth]{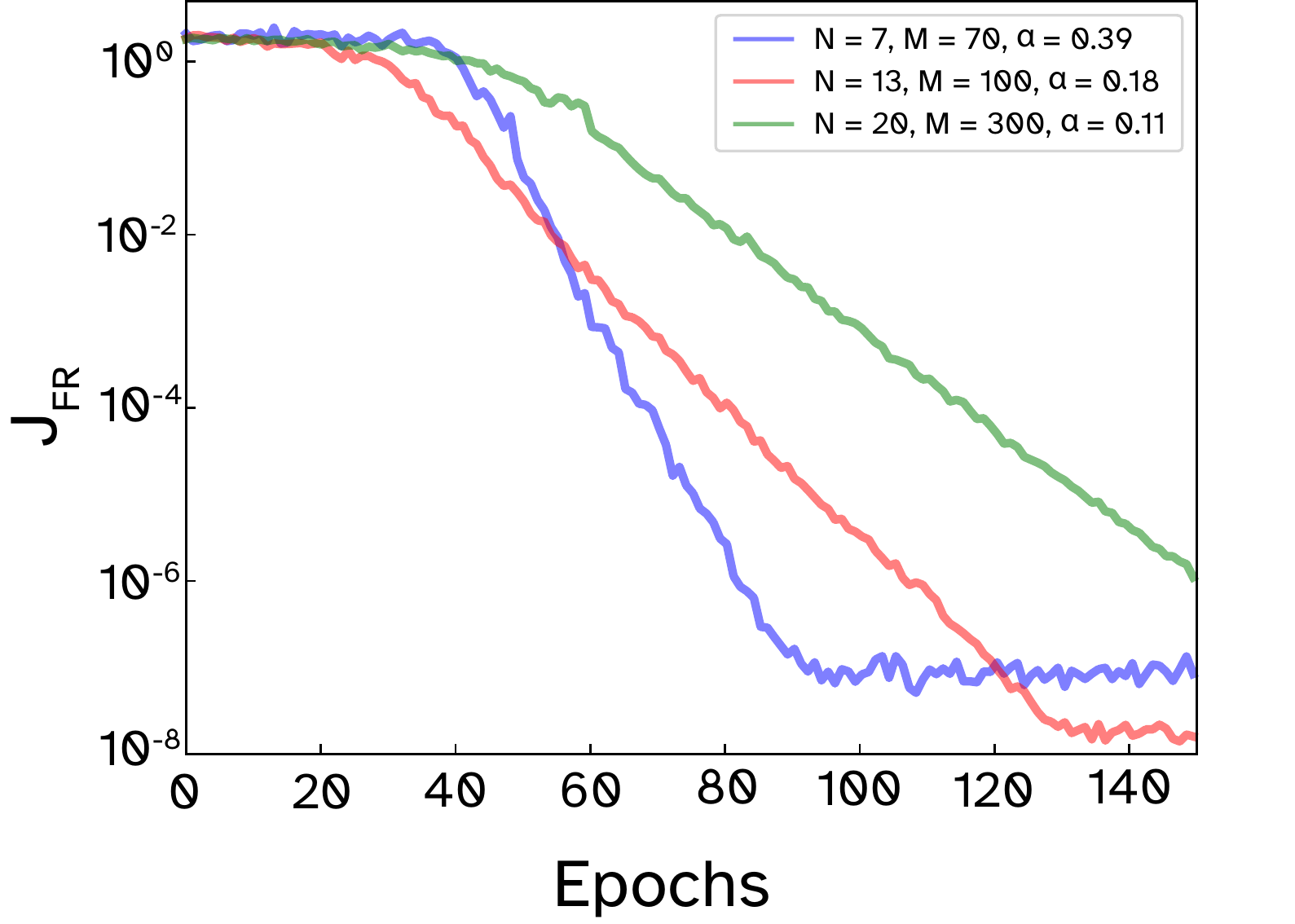}} a) \\
\end{minipage}
\hfill
\begin{minipage}[h]{0.48\linewidth}
\center{\includegraphics[width=1.11\linewidth]{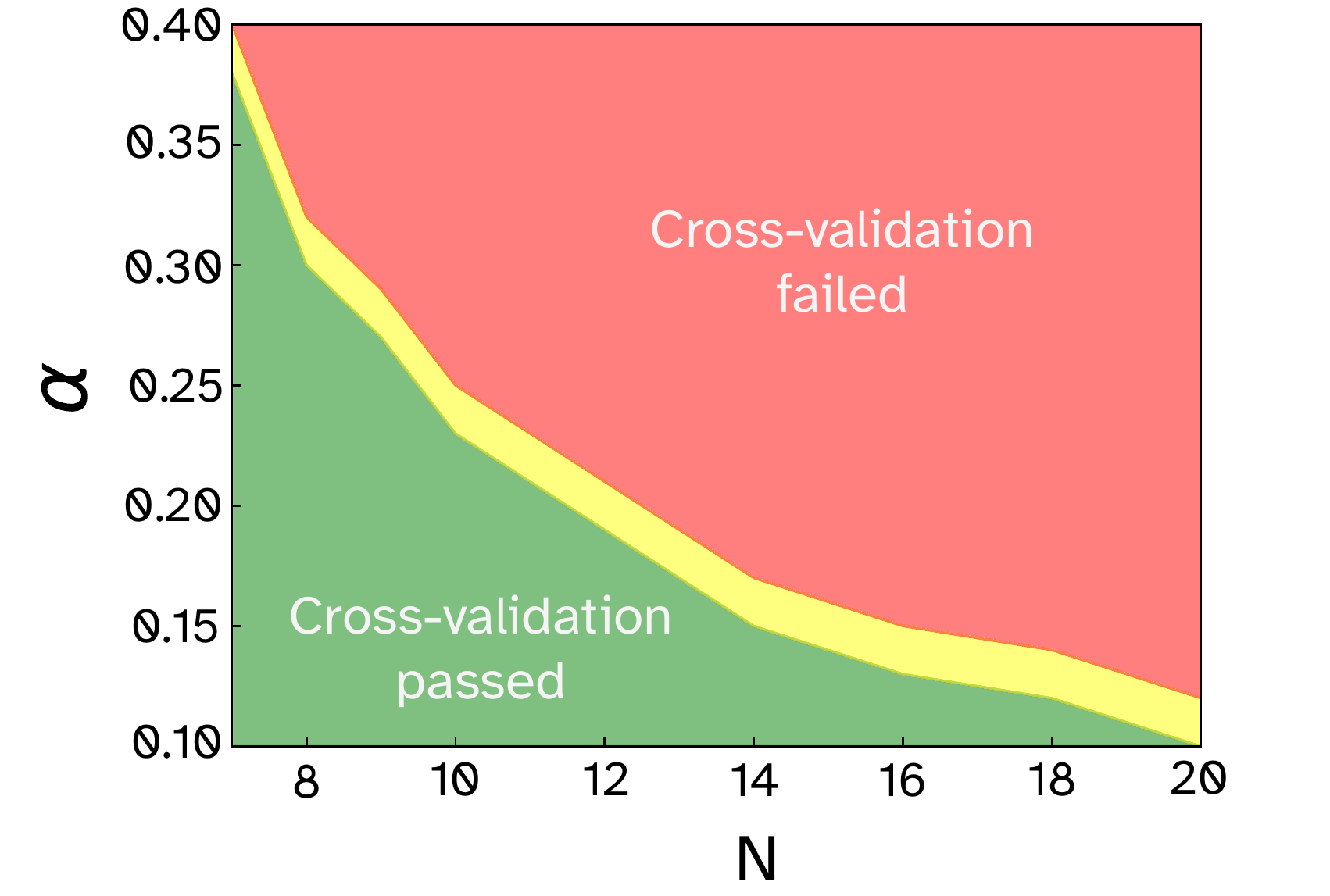}} b) \\
\end{minipage}
\caption{Investigation of the training process using a priori knowledge about basis matrices $U_{\ell}$. a) - Examples of training using a priori knowledge of basis matrices for large values of the number of optical modes $N$ of the interferometer, b) - Dependence of the maximum required $\alpha$ value, at which the model will be trained on the number of optical modes $N$.}
\label{Apriory}
\end{figure}

We use estimated matrices $U_{\ell}^{est}$ as the initial guess for our optimization routine. These matrices are chosen to be close to the ideal basis matrices $U_{\ell}$ used for training set generation. We get the initial guess using the procedure described be Eq. \ref{eq:deviation}.  The Fig. \ref{Apriory} shows the convergence of the learning algorithm employing the knowledge of the basis matrix unitary up to a certain precision. The {\it a priori} information about the basis matrices enabled learning the model of the interferometer up to $N=20$ using the same computational hardware. The larger the interferometer dimension $N$ the higher the precision of the $U_{\ell}^{est}$ estimation must be. The Fig. \ref{Apriory}b) illustrates the regions of the $\alpha$ value depending on the size of the interferometer where the algorithm still accurately learns the model $\mathcal{M}$. 

\section{Discussion}\label{sec:discussion}

The demonstrated approach to interferometer programming stands out with several major advantages. First and foremost the method is agnostic of the architecture of the interferometer. Any universal interferometers reported in literature \cite{Reck1994, Clements2016, Saygin2020, Fldzhyan2020} admit some form of the expansion Eq. \ref{Expansion} - the optical mode mixing elements interleaved with phase shifters. This means that both the gist of the algorithm and the mathematical framework fit any architecture of choice. This assumptions remains valid unless the mode mixers and the phase shifters are considered as the independent elements. Next, the output of the learning algorithm is the complete interferometer model taking into account the transformation of the mode mixing elements in the fabricated device. The model answers the question how close the required unitary $U_{0}$ can be approximated by the specific device under study and can pinpoint ares of unitary space inaccessible for the device due to design restrictions or the fabrication flaw. This feature has to be compared with typical calibration data used for programming the interferometer based on MZI blocks. The phase shifters are calibrated but no knowledge is available about the real transformation of the beamsplitters comprising the MZI. This fact leads to the necessity of running optimization procedure to improve the fidelity of the implemented unitary if some of the beamsplitters don't meet the designed transmission coefficients. Lastly the presented algorithm is essentially the reconstruction routine for the inner fixed optical elements of the complex interferometric circuit. Hence it can be adopted for probing the quality of the optical subcircuits located inside the larger optical scheme.

The bottlenecks of the proposed algorithm are related to the experimental issues. The $J_{FR}$ Frobenius metric requires exact measurement of the unitary elements' modulus and phase. Several reconstruction methods have been proposed and verified \cite{Laing2012, RahimiKeshari2013, Tillmann_2016, Spagnolo2017, Suess2020}. Some of them \cite{Laing2012, Suess2020} provide only partial information about the transformation matrix of the interferometer omitting phases which are impossible to reconstruct using the method-specific dataset. Any method will inevitably suffer from the path-dependent optical loss which is impossible to distinguish and attribute to the particular path inside the circuit. Another issue which is not covered by our algorithm arises from the crosstalks between the phase shifters. Our framework assumes that the phases in different paths are enabled independently which is not the case due to the crosstalks between different phase modulating elements. Luckily the integrated photonic modulator implementations typically exhibit extremely low crosstalks \cite{Zhang2020, Jiang2018}.

We believe that our results will enable opportunities to employ new programmable optical interferometer architectures for both classical and quantum applications. 

\section{Acknowledgements}
This work was supported by Russian Foundation for Basic Research
grant No 19-52-80034 and by Russian Science Foundation (RSF), project No: 19-72-10069. I. V. Dyakonov acknowledges support by Innopraktika.

%\tableofcontents

\bibliography{article}

\onecolumngrid

\section*{Supplemental material}

\appendix
\section{Computing gradients for the learning algorithm}\label{app:gradients}

According to the expressions (\ref{Expansion}) and (\ref{eq:Aux_matrix1}), $U = A_{\ell} U_{\ell} B_{\ell}$. Hence, we have:

\begin{equation}
u_{ij} = \sum_{k = 1}^N \sum_{m = 1}^N a^{(\ell)}_{ik} u^{(\ell)}_{km} b^{(\ell)}_{mj}.
\end{equation}

We calculate the derivative:

\begin{equation}\label{Grad_u}
\dfrac{\partial u_{ij}}{\partial u^{(\ell)}_{pt}} = \sum_{k = 1}^N \sum_{m = 1}^N a^{(\ell)}_{ik} \dfrac{\partial u^{(\ell)}_{km}}{\partial u^{(\ell)}_{pt}} b^{(\ell)}_{mj} = \sum_{k = 1}^N \sum_{m = 1}^N a^{(\ell)}_{ik} \delta_{kp} \delta_{mt} b^{(\ell)}_{mj} = a^{(\ell)}_{ip} b^{(\ell)}_{tj}.
\end{equation}

Let the unitary elements be $u_{ij}^{({\ell})} = x_{ij}^{({\ell})} + i y_{ij}^{({\ell})}$and we want to calculate the derivatives of the complex elements $u_{ij}$ of the matrix $U$ with respect to the real $x^{(\ell)}_{pt}$ and the imaginary $y^{(\ell)}_{pt}$  parts of the complex number $u_{ij}$:

\begin{equation}
\dfrac{\partial u_{ij}}{\partial x^{(\ell)}_{pt}} = \dfrac{\partial u_{ij}}{\partial u^{(\ell)}_{pt}} \dfrac{\partial u^{(\ell)}_{ij}}{\partial x^{(\ell)}_{pt}} = \dfrac{\partial u_{ij}}{\partial u^{(\ell)}_{pt}} = a^{(\ell)}_{ip} b^{(\ell)}_{tj} \hspace{0.3cm} \text{and} \hspace{0.3cm} \dfrac{\partial u_{ij}}{\partial y^{(\ell)}_{pt}} = \dfrac{\partial u_{ij}}{\partial u^{(\ell)}_{pt}} \dfrac{\partial u^{(\ell)}_{ij}}{\partial y^{(\ell)}_{pt}} = i \dfrac{\partial u_{ij}}{\partial u^{(\ell)}_{pt}} = i a^{(\ell)}_{ip} b^{(\ell)}_{tj}.
\end{equation}

We convert the obtained expressions to the matrix form. Using the auxiliary matrices $\Delta^{(mn)}$, which elements are zeros, except $\Delta^{(mn)}_{mn}=1$, we can transform the formula (\ref{Grad_u}) to take the following form:

\begin{equation}
\dfrac{\partial u_{ij}}{\partial u^{(\ell)}_{pt}} = a^{(\ell)}_{ip} b^{(\ell)}_{tj} = \sum_{k = 1}^N \sum_{m = 1}^N a^{(\ell)}_{ik} \Delta^{(pt)}_{km} b^{(\ell)}_{mj} \hspace{0.3cm} \Rightarrow \hspace{0.3cm} \dfrac{\partial U}{\partial u_{ij}^{({\ell})}} = A_{\ell} \Delta^{(ij)} B_{\ell}.
\end{equation}

Finally we arrive to the expression:

\begin{equation}
\boxed{
\begin{aligned}
\hspace{0.3cm}\dfrac{\partial U}{\partial x_{ij}^{({\ell})}} = A_{\ell} \Delta^{(ij)} B_{\ell}, \hspace{0.3cm} \\
\hspace{0.3cm}\dfrac{\partial U}{\partial y_{ij}^{({\ell})}} = i A_{\ell} \Delta^{(ij)} B_{\ell}. \hspace{0.3cm}
\end{aligned}
}
\end{equation}

\section{Computing gradients for the tuning task}

According to the expressions (\ref{Expansion}) and (\ref{eq:Aux_matrix2}), $U = C_{\ell} \Phi_{\ell} D_{\ell}$. Hence, we have:

\begin{equation}
u_{ij} = \sum_{k = 1}^N c^{(\ell)}_{ik} \left(\sum_{m = 1}^N e^{i \varphi_{lk}} \delta_{km} d^{(\ell)}_{mj}\right) = \sum_{k = 1}^N c^{(\ell)}_{ik} e^{i \varphi_{lk}} d^{(\ell)}_{kj}
\end{equation}

We calculate the derivative:

\begin{equation}
\begin{split}
\dfrac{\partial u_{ij}}{\partial \varphi_{pt}} &= \dfrac{\partial}{\partial \varphi_{pt}} \left(\sum_{k = 1}^N c^{(\ell)}_{ik} e^{i \varphi_{\ell k}} d^{(\ell)}_{kj}\right) = \sum_{k = 1}^N c^{(\ell)}_{ik} \left(\dfrac{\partial}{\partial \varphi_{pt}} e^{i \varphi_{\ell k}}\right) d^{(\ell)}_{kj} = \\
&= \sum_{k = 1}^N c^{(\ell)}_{ik} i e^{i \varphi_{\ell k}} \dfrac{\partial \varphi_{\ell k}}{\partial \varphi_{pt}} d^{(\ell)}_{kj} = \sum_{k = 1}^N i c^{(\ell)}_{ik} e^{i \varphi_{\ell k}} \delta_{\ell p} \delta_{kt} d^{(\ell)}_{kj} = i e^{i \varphi_{\ell k}} c^{(\ell)}_{ik} d^{(\ell)}_{kj} \delta_{\ell p}
\end{split}
\end{equation}

We get the expression for $\dfrac{\partial u_{ij}}{\partial \varphi_{\ell k}}$:

\begin{equation}\label{Grad_phi}
\dfrac{\partial u_{ij}}{\partial \varphi_{\ell k}} = i e^{i \varphi_{\ell k}} c^{(\ell)}_{ik} d^{(\ell)}_{kj}
\end{equation}

We will now express \ref{Grad_phi} in matrix form.We introduce auxiliary matrices $\Delta^{(kk)}$, which elements are all zeros, except $\Delta^{(kk)}_{kk}=1$. Then the formula (\ref{Grad_phi}) transforms to:

\begin{equation}
\dfrac{\partial u_{ij}}{\partial \varphi_{\ell k}} = i e^{i \varphi_{\ell k}} c^{(\ell)}_{ik} d^{(\ell)}_{kj} = i e^{i \varphi_{\ell k}} \sum_{j = 1}^N \sum_{m = 1}^N c^{(\ell)}_{ij} \Delta^{(kk)}_{jm} d^{(\ell)}_{mj}
\end{equation}
Then we have:
\begin{equation}
\boxed{
\hspace{0.3cm}\dfrac{\partial U}{\partial \varphi_{\ell k}} = i e^{i \varphi_{\ell k}} C_\ell \Delta^{(kk)} D_\ell. \hspace{0.3cm}
}
\end{equation}

\section{The deviation calibration}\label{app:deviation}

Often in quantum technologies and when working with unitary matrices, the {\it fidelity} value is used, which determines the degree of coincidence of two unitary matrices $U$ and $\bar{U}$. In our work, we define it as follows:

\begin{equation}
F(U, \bar{U}) \equiv \dfrac{1}{N^2} | Tr(\bar{U}^{\dagger} U) |^2 
\end{equation}

\begin{figure}[h!]
\center{\includegraphics[width=0.75\linewidth]{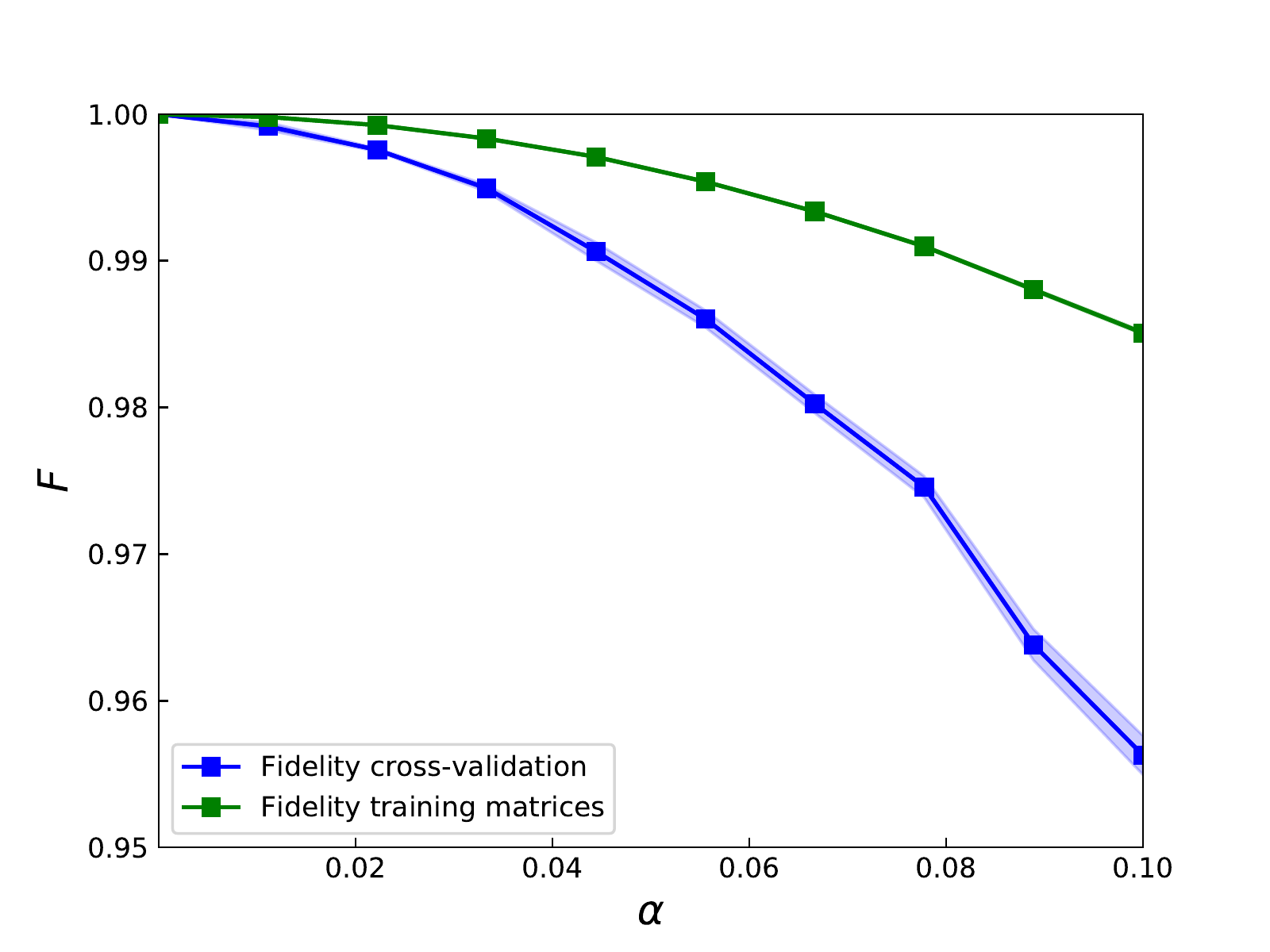}} \\
\caption{Dependence of the mean fidelity (Averaging over 1000 unitary matrices) with which the trained model can produce unitary matrices for the given phases (blue) and the mean fidelity (averaging over 10000 unitary matrices) with which the matrices from the training set (green) are known on the $\alpha$ parameter for size matrix $N = 2$.}
\label{Apriory_analytics}
\end{figure}

\end{document}